\documentclass[journal]{IEEEtran}
\usepackage[subpreambles=false]{standalone}
\usepackage{ieee-tsp2020}
\usepackage{import}
\usepackage{eso-pic}

\graphicspath{{img/}{sections/img/}}

\begin{document}
	\title{Online Joint State Inference and Learning of Partially Unknown State-Space Models}
	\author{Anton~Kullberg, Isaac~Skog,~\IEEEmembership{Senior Member,~IEEE}, Gustaf~Hendeby,~\IEEEmembership{Senior Member,~IEEE}%
		\thanks{\noindent A. Kullberg, I. Skog and G. Hendeby are with the Department of Electrical Engineering, Division of Automatic Control, Linköping University, Linköping, SE-58183, Sweden (E-mail: \{anton.kullberg, isaac.skog, gustaf.hendeby\}@liu.se).}%
		\thanks{This paper has supplementary downloadable material available at https://ieeexplore.ieee.org, provided by the author. The material includes a derivation of the motion model used in the second numerical example. This material is $\SI{8}{\mega\byte}$ in size.}
		\thanks{This work was partially supported by the Wallenberg AI,
			Autonomous Systems and Software Program (\textsc{WASP}) funded
			by the Knut and Alice Wallenberg Foundation.}
	}
\IEEEpubid{}
\markboth{}{}
\AddToShipoutPictureBG*{%
	\put(0,20){
		\hspace*{\dimexpr0.075\paperwidth\relax}
		\parbox{.84\paperwidth}{\footnotesize DOI: 10.1109/TSP.2021.3095709~\copyright2021 IEEE. Personal use of this material is permitted. Permission from IEEE must be obtained for all other uses, in any current or future media, including reprinting/republishing this material for advertising or promotional purposes, creating new collective works, for resale or redistribution to servers or lists, or reuse of any copyrighted component of this work in other works.}%
}}
\maketitle

\begin{abstract}
A computationally efficient method for online joint state inference and dynamical model learning is presented. The dynamical model combines an a priori known, physically derived, state-space model with a radial basis function expansion representing unknown system dynamics and inherits properties from both physical and data-driven modeling. 
The method uses an extended Kalman filter approach to jointly estimate the state of the system and learn the unknown system dynamics, via the parameters of the basis function expansion. The key contribution is a computational complexity reduction compared to a similar approach with globally supported basis functions. By using compactly supported radial basis functions and an approximate Kalman gain, the computational complexity is considerably reduced and is essentially determined by the support of the basis functions. The approximation works well when the system dynamics exhibit limited correlation between points well separated in the state-space domain. The method is exemplified via two intelligent vehicle applications where it is shown to: (i) have competitive system dynamics estimation performance compared to the globally supported basis function method, and (ii) be real-time applicable to problems with a large-scale state-space.
\end{abstract}


\section{Introduction}

\IEEEPARstart{S}{tate} estimation is a fundamental problem in many areas, such as robotics, target tracking, economics, etc. For a general nonlinear state-space model, typical techniques used in state estimation include Bayesian filters such as the \emph{extended Kalman filter} (\abbrEKF), \emph{unscented Kalman filter} (\abbrUKF) or the \emph{particle filter} (\abbrPF) \cite{Sarkka2010:BayesianFiltering}. Fundamental to these techniques is the specification of a \emph{state-space model} (\abbrSSM), which describes the dynamics of the underlying system and how the state of the system relates to the observations. These models can be broadly classified into three different categories: white-box (physically derived), gray-box (semi-physical) and black-box (non-physical) models. Black-box models generally do not assume any prior knowledge of the system. One particular issue with black-box models is that their generalization properties outside training regions is not known. Thus, black-box models may not be suited for operation critical applications. On the other hand, white-box modeling use domain-specific knowledge to derive a model based on the physical properties of the system. For instance, in the case of vehicle tracking, constant velocity or coordinated turn models are commonly used as first-order physical models to describe the vehicle dynamics \cite{Li2003}. These first-order physical models summarize any unknown behavior as white process noise, which leads to a state estimation bias \cite{Ljung1998}.One way to overcome the problems with white-box and black-box models is to use a gray-box modeling approach and combine the first-order physical model with a generic black-box model and try to learn (identify) a more refined model on-the-fly. Such a model is nearly as flexible as black-box models, yet can still guarantee minimum performance outside training regions when used for state estimation.

One way of constructing such a gray-box model is to augment a physically derived \abbrSSM with a black-box model structure. The black-box model augmentation can be chosen in a variety of ways; for instance, as a \emph{Gaussian process} (\abbrGP) \cite{Veiback2019:EKFGP, Kullberg2020}, as a basis function (\abbrBF) expansion \cite{Svensson2015}, or as a neural network. Henceforth, we shall refer to these models as {\abbrSSM}+\abbrGP, {\abbrSSM}+\abbrBF, etc. Another, highly related gray-box model, is the \emph{Gaussian process state-space model} (\abbrGPSSM), which was recently developed for joint state inference and system identification \cite{Turner2010:SSInferencewithGP, Frigola2013, Ko2009, Liu2021}. In the \abbrGPSSM, both the state dynamics and observation model are viewed as separate {\abbrGP}s and are learned jointly. Further, prior system information can be included in the state-space model as a known term \cite{Svensson2017, Berntorp2019a, Berntorp2021}.

The use of a \abbrGP model is attractive because of its non-parametric nature. However, since it uses all of the accumulated data during learning and inference, it is in practice still necessary to parameterize it to be able to use it efficiently. Particularly, the computational complexity of the standard \abbrGP is $\bigO(N^3)$ and $\bigO(N^2)$ during learning and inference, respectively, where $N$ is the number of data points. Further, the storage complexity is $\bigO(N^2)$. One way of reducing the complexity is to split the \abbrGP input into smaller, assumed independent, regions and fit a separate \abbrGP per region, such as in \cite{Lederer2020, Kok2018}. In those cases, it becomes important to handle the boundaries of the regions and it is not completely clear how to choose the shape and size of the regions. Another alternative is to parametrize the \abbrGP with inducing points, see \emph{e.g.}, \cite{Quinonero-Candela2005:SparseGPregression,Snelson2005:SPGP}. This is a way of summarising the accumulated function observations in a set of discrete points, thereby bounding the computational complexity during learning and inference. The computational complexity then drops to $\bigO(NM^2)$ and $\bigO(M^2)$ for learning and inference, respectively, where $M$ is the number of inducing points. Further, the storage complexity is reduced to $\bigO(MN+ M^2)$. Hence, to see any major benefit, the number of inducing points, $M$ should be much smaller than $N$ \cite{Quinonero-Candela2005:SparseGPregression}. Essentially all of \cite{Turner2010:SSInferencewithGP, Frigola2013, Ko2009, Liu2021} use inducing point approaches to reduce the computational burden of the \abbrGP.

A related approach is the {\abbrSSM}+\abbrBF model, which instead parametrizes the unknown function using a \abbrBF expansion. In the limit, \emph{i.e.}, when the number of basis functions goes to infinity, these two function representations are equivalent, under certain conditions, and share many properties. Particularly, \cite{Svensson2017, Berntorp2019a, Berntorp2021} use a regularized \abbrBF expansion that can be viewed as a {\abbrSSM}+\abbrGP model. The benefit of using a \abbrBF expansion is that it offers an intuitive appeal with regard to function approximation and is easy to include in the state-space model. However, in its standard formulation, it still suffers from the same computational issues as the {\abbrSSM}+\abbrGP methods. Recently, \cite{Berntorp2021} extended the \abbrGPSSM inference framework in \cite{Svensson2017} to an online setting and used the method to estimate lateral tire-friction in real-time. However, \cite{Berntorp2021} was focused on limited state-spaces, such that the function expansion requires only a limited amount of basis functions $(< 100)$. The common limitation of the standard {\abbrSSM}+\abbrBF and {\abbrSSM}+\abbrGP models, is that all of the parameters of the \abbrGP or \abbrBF expansion are required for each prediction/update step of the filtering algorithms that are applied to them. This effectively limits the methods to problems where the number of inducing points/basis functions can be kept less than a few thousand. Hence, they are not suitable when the bandwidth or the support of the function to be learned grows.

\IEEEpubidadjcol 

Therefore, in this paper, a scalable online method for joint state inference and the system identification of {\abbrSSM}+\abbrBF gray-box models is presented. By using basis functions with a limited support the method is able to handle models with large-scale state-spaces. The connection between the suggested {\abbrSSM}+\abbrBF model and the {\abbrSSM}+\abbrGP model is highlighted. Further, the computational complexity of performing state inference and model learning with the suggested {\abbrSSM}+\abbrBF model using an extended Kalman filter is analyzed. The performance of the suggested {\abbrSSM}+\abbrBF model and inference framework is demonstrated using two intelligent vehicle applications.

\label{sec:introduction}

\section{Problem Formulation}

A general gray-box \abbrSSM with partially unknown state dynamics is given by
\begin{subequations}
	\label{eq:int:generalmodel}
	\begin{align}
	\state_{k+1} &= \vec{f}_k\Big(\state_{k},\inp_{k},\vec{w}_{k}, \bunf_k(\state_{k}, \inp_k)\Big)\\
	\obs_k &= \vec{h}_k(\state_k, \inp_k, \vec{e}_k).
	\end{align}
\end{subequations}
Here, $\state_k$ and $\obs_k$ denotes the unknown system state and observations at time $k$, respectively. Further, $\inp_k$ is a known input to the system. Moreover, $\vec{w}_k$ and $\vec{e}_k$ denote the process and observation noise, respectively. The process and observation noises are assumed to be mutually independent white noise processes with covariance matrices $\vec{Q}_k$ and $\vec{R}_k$, respectively. The functions $\bknf_k$ and $\vec{h}_k$, which specify the state dynamics and the observation model, are chosen a priori, and are here assumed to be differentiable. Lastly, $\bunf_k$ is an unknown function, dependent on the state $\state_k$ and the known input $\inp_k$, which is to be inferred from data. Further, depending on how the function $\bknf_k$ is chosen, $\bunf_k$ can be interpreted in different ways. For instance, if $\bknf_k$ is chosen as a \emph{constant velocity} (\abbrCV) model \cite{Li2003}, $\bunf_k$ represents the system acceleration, see \cref{subsec:experiments:constantvel}. Moreover, note that $\bunf_k$ could be additive or multiplicative and could affect but a subset of $\state_k$. 

Data from a single system allows the identification (learning) of the unknown function $\bunf_k$. However, data from multiple systems, whose dynamics depend on $\bunf_k$, is often available. For instance, ships sailing in certain regions are all influenced by the same currents, neighboring buildings are all affected by the same winds, and road vehicles traversing the same road network are all subject to the same road conditions. Hence, a particularly interesting application of the model \cref{eq:int:generalmodel} is when collaborative learning of $\bunf_k$ is possible. It is also worth mentioning that, within this paper, explicit references to several systems have been omitted for notational brevity.

\subsection{Basis Function Expansion}
As previously mentioned, the function $\bunf_k$ can be modeled in a variety of ways.
Here, it is modeled as a basis function expansion, \emph{i.e.}, 
\begin{equation}
\label{eq:int:basfunexpansion}
\left[\bunf_k\right]_{j} =  (\bbasfun_k^j)^\top\bweight^j_k\quad j=1,\dots,J%
\end{equation}
where $\left[\bunf_k\right]_j$ denotes the $j$:th component of $\bunf_k$. Further, 
\begin{align*}
\bbasfun_k^j&=\begin{pmatrix}\basfun^j_1 &\dots& \basfun^j_{n_\weight^j }\end{pmatrix}^\top,\\
\bweight^j_k&=\begin{pmatrix}\weight^j_{k,1} &\dots &\weight^j_{k,n_\weight^j}\end{pmatrix}^\top,
\end{align*}
where $\basfun^j_i=\basfun^j_i(\state_k,\inp_k)$ is the $i$:th basis function corresponding to component $j$ of $\bunf_k$ and $\weight^j_{k,i}$ is the weight of this basis function at time $k$. Moreover, the dependence of $\bunf_k$ on $\state_k$ and $\inp_k$ is solely through the basis functions $\basfun^j_i$. Lastly, $n^j_\weight$ denotes the number of basis functions used to represent component $j$ of $\bunf_k$.

The basis function expansion \cref{eq:int:basfunexpansion} is closely related to the \abbrGP representation of $\bunf_k$. With a Gaussian prior on the weights, $\bweight^j_0\sim\Ndist(0,\sigma_j^2I)$, $\left[\bunf_k\right]_j$ will correspond to a \abbrGP with kernel
\begin{equation}
\label{eq:int:gpkernelfrombasfun}
\kappa^j(\state,\state') = \sigma_j^2\sum_{i=1}^{n_\weight^j}\basfun^j_i(\state)\basfun^j_i(\state'),
\end{equation}
if the basis functions $\basfun^j_i$ are such that the matrix
\begin{equation}
\vec{K}^j=\sigma_j^2\begin{bmatrix}\basfun^j_1(\state)\basfun^j_1(\state') & \dots & \basfun^j_1(\state)\basfun^j_{n_\weight^j}(\state')\\
\vdots & \ddots & \vdots\\
\basfun^j_{n_\weight^j}(\state)\basfun^j_1(\state') & \dots & \basfun^j_{n_\weight^j}(\state)\basfun^j_{n_\weight^j}(\state')
\end{bmatrix}
\end{equation}
is at least positive semi-definite. Here, $\state$ and $\state'$ are two data points, see, \emph{e.g.}, \cite{RasmussenWilliams2006:GPforML}. Particularly, if the basis function is chosen as a Gaussian, \emph{i.e.},
\begin{equation}
\label{eq:prob:gaussianbasfun}
\basfun^j_i(\state)=\exp\left(-\frac{(\state-\bfcenter^j_i)^2}{2l^2}\right),
\end{equation}
with center $\bfcenter^j_i\in\Bfcenter$, where $\Bfcenter$ is the set of all $\bfcenter^j_i$, and scaling parameter $l$. Then, if $\Bfcenter$ is a regular, equally-spaced grid, the basis function expansion will correspond to a \abbrGP with a standard squared exponential kernel as the number of basis functions $n^j_\weight\to\infty$ \cite{RasmussenWilliams2006:GPforML}.

The basis function expansion can, in other words, be seen as a parametrization of a \abbrGP. The main difference between the approaches is what is being \q{learned.} In an inducing point approach, function values are learned whereas in a basis function expansion, basis function weights are learned. Regardless of the semantic differences, both of the approaches suffer from the same problems, which primarily concern a restrictive computational complexity. Henceforth, we shall focus on the basis function expansion \eqref{eq:int:basfunexpansion} and begin with casting the problem of learning (identifying) the function $\bunf_k$ into a state estimation problem. Thereafter, we will present a method to reduce the computational complexity.

To that end, form the augmented state vector and process noise vector
\begin{equation}
	\state^e_k=\begin{pmatrix}\state_k^\top & \bweight_k^\top\end{pmatrix}^\top\quad 
	\vec{w}^e_k =\begin{pmatrix}
	\vec{w}_k^\top & \bnoise_k^\top
	\end{pmatrix}^\top,
\end{equation}
where 
\begin{equation*}
\bweight_k= \begin{pmatrix}(\bweight_k^1)^\top & \cdots & (\bweight_k^J)^\top\end{pmatrix}^\top.
\end{equation*}
Then, the model described by \cref{eq:int:generalmodel,eq:int:basfunexpansion} can be written as
\begin{subequations}
\label{eq:int:filtproblem}
\begin{align}
\state^e_{k+1} &= \bknf^e_k(\state^e_k, \inp_k, \vec{w}_k^e)\\
\bknf^e_k &\triangleq \begin{pmatrix}
\bknf_k(\state_k, \inp_k, \vec{w}_k, \Basfun_k\bweight_k)\\
\bweight_k+\bnoise_k
\end{pmatrix},
\end{align}
\end{subequations}
where 
\begin{equation*}
\Basfun_k =\diag\begin{pmatrix}
(\bbasfun^1_k)^\top & \cdots & (\bbasfun^J_k)^\top
\end{pmatrix},
\end{equation*}
and $\bnoise_k$ is a white process noise with covariance matrix $\bm\Sigma_k$, that allows $\bunf_k$ to change over time and is assumed to be independent of $\vec{w}_k$ and $\vec{e}_k$. Hence, the problem of learning the function $\bunf_k$ has been cast into a state estimation problem, and both the states $\state_k$ and function $\bunf_k$ can be inferred simultaneously.

\subsection{Computational Complexity}\label{subsec:problemformulation:compcomplexity}
Focusing on the model defined by \cref{eq:int:generalmodel,eq:int:filtproblem}, the computational complexity of standard state estimation algorithms is in the order of $\bigO(d^3)$ or more \cite{Daum2005}, where $d$ is the dimension of the state vector. Hence, nothing has yet been gained in terms of computational complexity, thus leaving the problem of handling models in which the function $\bunf_k$ varies quickly or has a large support, unsolved. 
To simplify the forthcoming discussion about how the computation complexity can be reduced, we will focus on estimating the states of \cref{eq:int:filtproblem} using an \abbrEKF. This because (i) the \abbrEKF is one of the most commonly used nonlinear filtering approaches, and (ii) the \abbrEKF is a computationally efficient approximation of the general recursive Bayesian filter, which fits well into the focus of computational efficiency here. Noteworthy is that the linearizations in the \abbrEKF may cause problems with highly nonlinear systems and particularly if the function $\bunf_k$ being learned is highly nonlinear in the states $\state_k$. In such cases, it may be necessary to substitute the proposed \abbrEKF for a marginalized \abbrPF, as in \cite{Berntorp2021, Svensson2017}, or some type of linear regression Kalman filter, such as the \abbrUKF \cite{Julier1997:UKF}, or the \abbrSTKF \cite{Steinbring2014}. Even though the complexity reduction method presented next is not directly applicable to these cases, large parts of it can be used.

The \abbrEKF propagates the first two moments in each iteration and hence, only the mean vector $\hat\state^e$ and covariance matrix $\vec{P}$, need to be stored, defined as
\begin{subequations}
\begin{align}
\hat\state_k^e&=\begin{bmatrix}\hat\state_k \\ \hat\bweight_k\end{bmatrix}=\mathbb{E}\begin{bmatrix}\state_k \\ \bweight_K\end{bmatrix}\\
\vec{P}_k &= \mathbb{E}[(\state_k^e-\hat\state_k^e)(\state_k^e-\hat\state_k^e)^\top].
\end{align}
\end{subequations}

If an \abbrEKF is directly designed based upon model \cref{eq:int:filtproblem}, the filter has a computational complexity of $\bigO((n_x^e)^3)$, where $n_x^e=\dim(\state^e_k)$.
However, using the fact that the weights $\bweight$ are modeled as a random walk process, the computational complexity can be reduced. By using this structure for the time update, the prediction mean is given by
\begin{subequations}
	\label{eq:ekf:timeupdate:mean}
	\gdef\kk{_{k|k}}
	\gdef\pkk{_{k+1|k}}
	\begin{align}
	\hat{\bunf}\kk &= \left(\vec{I}\otimes\bbasfun_k^\top(\hat\state\kk)\right)\hat{\bweight}\kk\triangleq\Basfun_k(\hat\state\kk)\hat\bweight\kk\\
	\hat{\state}\pkk &= \bknf_k(\hat{\state}\kk, \inp_{k+1}, \vec{w}_{k+1},\hat{\bunf}\kk)\\
	\hat{\bweight}\pkk &= \hat{\bweight}\kk.
	\end{align}
\end{subequations}
Further, the predictive covariance is given by
\begin{align}
\vec{P}\pkk=&\begin{bmatrix}
\Px\pkk & \Pxw\pkk\\
\Pwx\pkk & \Pw\pkk
\end{bmatrix} = \vec{F}_k\vec{P}\kk\vec{F}_k^*+\vec{Q}_k^{e}\nonumber\\
=&\begin{bmatrix}
\Fx & \Fw\\
\vec{0} & \vec{I}
\end{bmatrix}\begin{bmatrix}
\Px\kk & \Pxw\kk\\
\Pwx\kk & \Pw\kk
\end{bmatrix}\begin{bmatrix}
{\Fx}^\top & \vec{0}\\ 
{\Fw}^\top & \vec{I}
\end{bmatrix}\nonumber\\
&{+}\:\begin{bmatrix}
\vec{Q}_k & 0\\
0 & \bm{\Sigma}_k
\end{bmatrix}
\end{align}
which can be written as
\begin{subequations}
	\label{eq:ekf:timeupdate:covariance}
	\begin{align}
	\Px\pkk={}&\Fx\Px\kk{\Fx}^\top+\Fx\Pxw\kk{\Fw}^\top\nonumber\\
	&{+}\:\Fw\Pwx\kk{\Fx}^\top+\Fw\Pw\kk{\Fw}^\top+\vec{Q}_k\label{eq:ekf:timeupdate:pxx}\\
	\Pxw\pkk={}&\Fx\Pxw\kk+\Fw\Pw\kk\\
	\Pwx\pkk={}&(\Pxw\pkk)^\top\\
	\Pw\pkk={}&\Pw\kk+\bm{\Sigma}_k.
	\end{align}
\end{subequations}
Here, 
\begin{equation}
\label{eq:fgrad}
\vec{F}_k\triangleq\nabla_{\state^e}\bknf_k^{e}= \begin{bmatrix}
\nabla_{\state}\bknf_k & \nabla_{\bweight}\bknf_k\\
0 & \vec{I}
\end{bmatrix}\triangleq \begin{bmatrix}
\Fx & \Fw\\
0 & \vec{I}
\end{bmatrix}.
\end{equation}
Inspecting \Crefrange{eq:ekf:timeupdate:mean}{eq:ekf:timeupdate:covariance}, the main computational burden in the time update is related to \cref{eq:ekf:timeupdate:covariance}, and more specifically, the update of $\Px$ in \cref{eq:ekf:timeupdate:pxx}, where the complexity of the product $\Fw\Pw{\Fw}^\top$ is in the order of $\bigO(n_x n_\weight^2+n_x^2n_\weight)<\bigO((n_x^e)^3)$. Further, after a measurement, the corrected mean is given by
\begin{subequations}
	\gdef\kkm{_{k|k-1}}
	\label{eq:ekf:measupdate:mean}
	\begin{align}
	\hat{\state}\kk ={}& 
	\hat{\state}\kkm+\Lx(\obs_k-\Hx\hat{\state}\kkm)\\
	\hat{\bweight}\kk ={}& \hat{\bweight}\kkm+\Lw(\obs_k-\Hx\hat{\state}\kkm),
	\end{align}
\end{subequations}
and the corresponding covariance update, in the less commonly used Joseph's form, is given by
\begin{equation}
\label{eq:ekf:measupdate:completejoseph}
\vec{P}\kk = (\vec{I}-\vec{K}_k\Hk)\vec{P}\kkm(\vec{I}-\vec{K}_k\Hk)^\top+\vec{K}_k\vec{R}\vec{K}_k^\top.
\end{equation}
Here,
\begin{subequations}
	\begin{align}
	\label{eq:hgrad}
	\Hk \triangleq{}& \nabla_{\state^e\kk}\vec{h}_k(\hat{\state}\kk, \inp_k, \vec{e}_k)=\begin{bmatrix}
	\Hx & \vec{0}
	\end{bmatrix}\\
	\vec{S}_k \triangleq{}& \vec{R}_k+\Hx\Px\kkm{\Hx}^\top\\
	\Lx \triangleq{}& \Px\kkm{\Hx}^\top{\vec{S}_k}^{-1}\\
	\Lw \triangleq{}&\Pwx\kkm{\Hx}^\top{\vec{S}_k}^{-1}\label{eq:kalmangaintheta},
	\end{align}
\end{subequations}
where \cref{eq:hgrad} follows from the independence between $\vec{h}_k$ and $\bweight_k$. The covariance update in \cref{eq:ekf:measupdate:completejoseph} can also be written in its separate components as
\begin{subequations}
	\label{eq:ekf:measupdate:covariance}
	\begin{align}
	\Px\kk ={}& (\vec{I}-\Lx \Hx)\Px\kkm(\vec{I}-\Lx\Hx)^\top+\Lx\vec{R}_k{\Lx}^\top\\
	\Pxw\kk ={}& (\vec{I}-\Lx \Hx)\Px\kkm(-\Lw\Hx)^\top\nonumber\\
	&{+}\:(\vec{I}-\Lx\Hx)\Pxw\kkm+\Lx\vec{R}_k{\Lw}^\top\\
	\Pwx\kk ={}& (\Pxw\kk)^\top\\
	\Pw\kk ={}& \Pw\kkm + \Lw\Hx\Px\kkm{\Hx}^\top{\Lw}^\top\nonumber\\
	&{-}\:\Lw\Hx\Pxw\kkm-\Pwx\kkm{\Hx}^\top{\Lw}^\top+\Lw\vec{R}_k{\Lw}^\top\label{eq:ekf:measupdate:costlyupdate}.
	\end{align}
\end{subequations}
Note that \cref{eq:ekf:measupdate:completejoseph,eq:ekf:measupdate:covariance} are guaranteed to be positive semi-definite regardless of the choice of gain $\vec{K}_k$, which shall be of importance later. 

From \Crefrange{eq:ekf:measupdate:mean}{eq:ekf:measupdate:covariance}, it is clear that the main computational complexity of the measurement update is related to \cref{eq:ekf:measupdate:costlyupdate}, which is in the order of $\bigO(n_\weight n_y^2+n_\weight^2n_y)$, where $n_y$ is the size of the observation vector $\obs$. Specifically, products involving both $\Lw$ and ${\Lw}^\top$ are the culprits, \emph{e.g.}, the product $\Lw\vec{R}_k{\Lw}^\top$. If it is (reasonably) assumed that $n_\weight\gg n_x$ and $n_\weight\gg n_y$, the computational complexity is thus approximately $\bigO(n_\weight^2)$. Even though this is a substantial reduction in computational complexity, it is still prohibitive if $n_\weight$ is large.

The main computational burden is caused by the basis functions $\basfun$. Assume that the basis functions have global support, \emph{i.e.}, $\basfun(\state)\neq 0,~\forall \state$, which could be, for instance, the Gaussian basis function \cref{eq:prob:gaussianbasfun}. Essentially, this means that regardless of $\state$, all of the basis functions will be non-zero and are required in both the time update and measurement update of the \abbrEKF. In particular, the matrices $\Pwx,\Pxw$ and $\Pw$ will be dense. The naive solution would be to decrease $n_\weight$ to yield a tractable complexity for the problem at hand. However, if the domain of $\bunf$ is large, then $n_\weight$ needs to be large to cover the entire domain. Similarly, if $\bunf$ varies quickly in regard to the input, then $n_\weight$ needs to be large enough to accurately represent the function. This leads to a conundrum as $n_\weight$ needs to be small enough to remain with a tractable algorithm, yet also large enough to represent complex functions. Hence, with a globally supported basis function, there is a trade-off between computational complexity and the types of functions $\bunf$ can represent.

To reduce the complexity of learning $\bunf$ and estimating the state $\state_k$, we seek a basis function $\basfun$ so that only a few of the basis function weights $\bweight$ need to be used at each filter step. In the following section, we present a basis function, which fulfills this criterion. We now restrict the basis functions $\basfun^j_i$ to the family of \emph{radial basis functions} ({\abbrRBF}s) \cite{Buhmann2003}, as they are universal approximators \cite{Park1991}. It does simplify the discussion, but we stress that it is possible to use other basis functions and reach similar conclusions. 

\label{sec:problemformulation}

\section{Computational Complexity Reduction Using Compact Basis Functions}\label{sec:computationalconsiderations}

The computational complexity is largely caused by the number of basis functions $n_\weight$. In particular, the problem is that, given a set of globally supported basis functions, all of the basis functions need to be evaluated in each time step, since $\state_k$ depends on the entire $\bweight_k$ (recall $\bunf_k=\Basfun_k\bweight_k$). If, instead, $\state_k$, at every time instant, only depended on a subset of $\bweight_k$, only a few weights would need to be updated in each step of the filtering algorithm. This restricts the possible choice of basis function to those with \emph{compact support}, \emph{i.e.}, basis functions which are only non-zero on a subset of $\mathcal{X}$, where $\state_k\in\mathcal{X}$. Let $\tilde{n}_\weight$ be the number of non-zero basis functions at any time $k$. Then, if $\tilde{n}_\weight\ll n_\weight$, the computational complexity can be greatly reduced.

\subsection{Basis Functions with Compact Support}\label{subsec:compactsupport}
In the family of {\abbrRBF}s, a number of basis functions have compact support \cite{RasmussenWilliams2006:GPforML, Buhmann2003}, \emph{i.e.},
\begin{equation}
\label{eq:bf:compactrequirement}
\exists c,\quad \text{s.t.}\quad \basfun^j_i(\state_k)=0,\quad \text{if}\quad \bfrad=\lVert\state_k-\bfcenter^j_i\rVert>c.
\end{equation}
These basis functions are commonly referred to as \emph{compactly supported radial basis functions} ({\abbrCSRBF}s). {\abbrCSRBF}s have been studied extensively in many different scientific disciplines, for instance, the geostatistical community \cite{Porcu2013, Furrer2006}. They have also been used in ensemble Kalman filters for atmospheric data assimilation \cite{Houtekamer2001, Hamill2001}. In the geostatistical community, {\abbrCSRBF}s have mainly been used for data interpolation, where the choice of basis function is of great importance in recovering the \q{true} function exactly. As alluded to before, in the context of this paper, the choice is important only for the interpretability as a \abbrGP and in general, the choice can be made more freely. Further, smoothness and continuous differentiability are convenient and sometimes necessary. For instance, if a standard \abbrEKF is used, differentiability is required. Hence, we shall restrict ourselves to Wendland functions \cite{Wendland1999}, which have been shown to be positive definite, up to some dimension $d$. Thus, with Wendland basis functions and a Gaussian prior on the weights, the basis function representation can be interpreted as a \abbrGP. Further, the Wendland functions have been shown to be of minimal degree, as well as being continuously differentiable both at the origin and the cut-off radius $c$. 

In the remainder of the paper, $\basfun$ is chosen as the particular Wendland function
\begin{equation}
\label{eq:basisfunction:crbf}
\basfun^j_i(\bfrad) = (1-\bfrad)^6_+(35(\bfrad)^2+18\bfrad+3)/3,
\end{equation}
where $(\cdot)_+=\max(0,~\cdot)$. Clearly, $\basfun^j_i(\bfrad)\equiv0$ if $\bfrad\geq1$. To get a larger or smaller support, $\bfrad$ can be scaled appropriately. Henceforth, $\bfrad$ is assumed to be scaled by $1/ \alpha$, which, for \emph{e.g.}, $\alpha=2$, increases the support by a factor two. Lastly, the \abbrCSRBF \cref{eq:basisfunction:crbf}, is positive definite in $d=2$ and $4$ times continuously differentiable \cite{Wendland1999}. 

By considering any function which satisfies \cref{eq:bf:compactrequirement}, the dynamics of $\state_k$ will now depend on just a few of the basis functions at every time instance $k$. Let 
$${\Ind^j_k(\state_k)\in\R^{\tilde{n}_\weight^j\times n_\weight^j}}$$ 
be a matrix-valued function that takes $\state_k$ as input and produces an indicator matrix such that only the \textit{active}, or non-zero, basis functions and weights remain (\emph{i.e.}, the products
$\Ind^j_k\bbasfun^j_k,~\Ind^j_k\bweight_k^j$ contain only non-zero basis functions and the corresponding weights). Moreover, note that the products produce a vector of size $\tilde{n}^j_\weight$. The expansion \cref{eq:int:basfunexpansion} can then be rewritten as
\begin{equation}
\label{eq:subsetbfexpansion}
\left[\bunf_k\right]_{j}= 
(\Ind^j_k\bbasfun^j_k)^\top \Ind^j_k\bweight^j_k=(\bbasfun^j_k)^\top(\Ind^j_k)^\top\Ind^j_k\bweight^j_k.
\end{equation}
Now, if $\tilde{n}^j_\weight\ll n^j_\weight$, the computational complexity is greatly reduced, since only these weights are needed in the recursions of the filter. Note that $\Ind^j_k$ is time-varying, as is $\tilde{n}^j_\weight$. Hence, the number of active basis functions varies over time, but is bounded from above. If the set of basis function centers $\Bfcenter$ is assumed to be a regular grid, the upper bound on the number of active basis functoins is completely determined by the density of $\Bfcenter$, as well as the support of the basis functions.

For an \abbrRBF expansion to be viable in large scale online applications, there are a few more things one needs to consider. First of all, the size and density of the grid of basis functions needs to be chosen in such a way that the entire state-space region of interest is covered. In the \abbrCSRBF case, one also needs to consider the scaling factor $\alpha$, \emph{i.e.}, the support of the basis functions. This is application-specific and the parameters can be found via maximum likelihood estimation, see \emph{e.g.} \cite{Nychka2015}, or they can be manually selected based on expert knowledge. Further, in the \abbrCSRBF case, one needs to consider how long spatial correlations to capture, as the support determines how long correlations that can be represented. 

In \cite{Nychka2015}, multiple grids with varying resolution and scaling parameters were considered to capture both short-range and long-range correlations. However, in a real-time scenario, the approach taken there is prohibitive, as all of the resolutions are active simultaneously, and thus, the computational complexity is still restrictive. Another multi-resolution basis function expansion was considered in \cite{Katzfuss2017}, where the multiple resolutions were instead used to vary the fineness of the grid. In this paper, we shall focus on manually selected grid resolutions and leave adaptivity or multi-resolution for future research. There are also a few more nuanced aspects to making the expansion real-time applicable, which we shall discuss next.

\begin{figure}[tbp]
	\centering
	\includegraphics[width=\columnwidth]{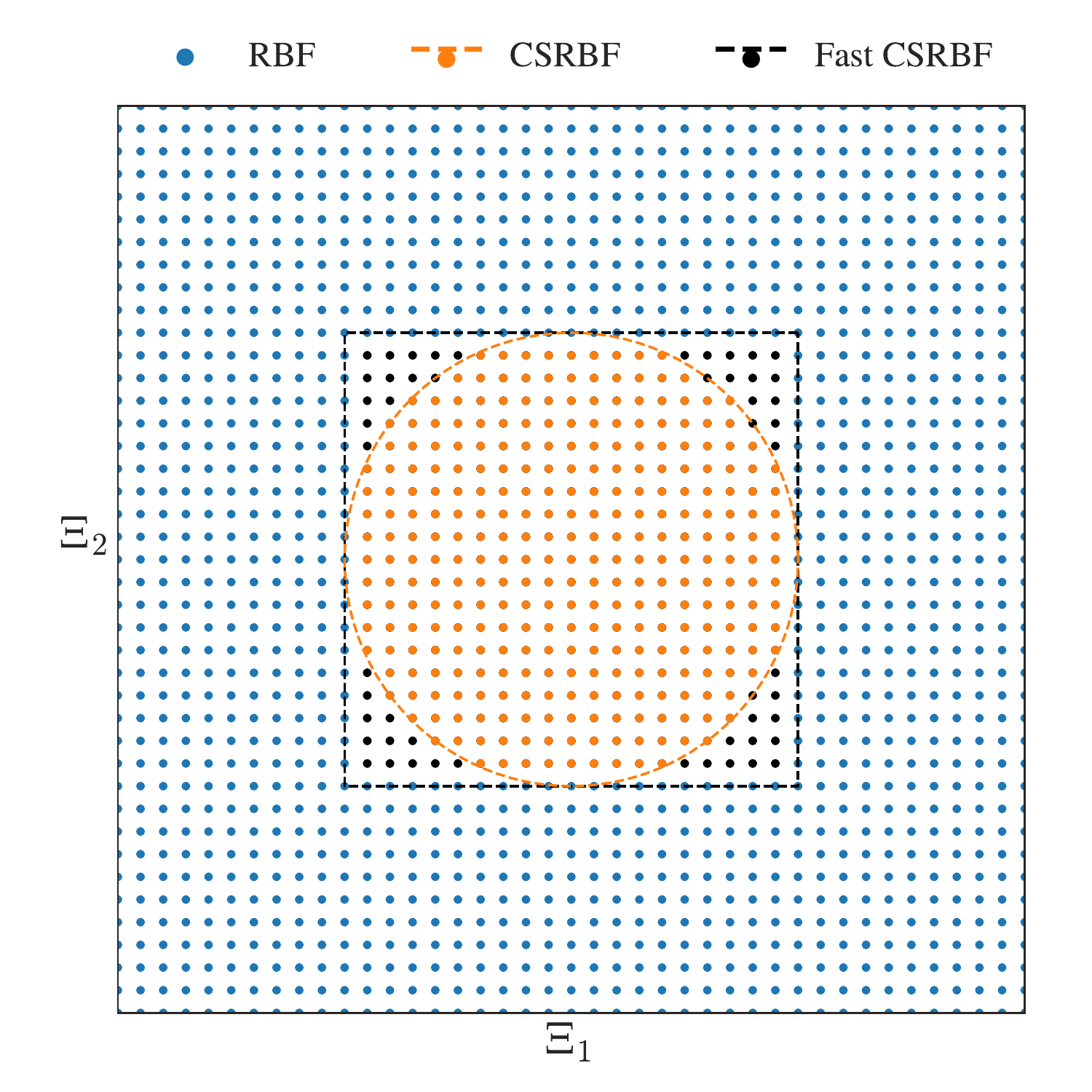}
	\caption{Basis functions used in the description of $\bunf_k(\state_k)$ when considering a grid ${\Bfcenter_1,\Bfcenter_2=[0,~4]}$ and $\state_k=[2,~2]^T$. Blue dots are for the global \abbrRBF description. The orange dots indicate the active (non-zero) basis functions in the \abbrCSRBF description of $\bunf_k(\state_k)$. The dots within the black square indicates the basis functions selected by the fast \abbrCSRBF method.}
	\label{fig:bfselection}
\end{figure}

\subsection{Finding Necessary Basis Functions}\label{subsec:findingbasisfunctions}
Assuming that the basis function expansion uses {\abbrRBF}s with compact support, how can $\Ind_k^j$ be determined? The naïve approach is to calculate the basis function values and remove any that are identically 0. If $n^j_\weight$ is large, this might prove to be computationally prohibitive. Instead, consider the special case when $\Bfcenter$ is a Cartesian grid (\emph{i.e.}, decomposes as ${\Bfcenter=\Bfcenter_1\times\Bfcenter_2\times\dots\times\Bfcenter_P}$, where $\Bfcenter_p$ are the center coordinates along the $p$:th dimension and $\times$ denotes the Cartesian product). Then, if the basis functions is a product basis function, \emph{i.e.},
\begin{equation}
\label{eq:bf:productbf}
\basfun_i^j(\state)=\prod_{p=1}^{P}\basfun_i^j([x]_p),
\end{equation}
where $[x]_p$ is the $p$:th element of $\state$, the basis function values can be computed in each separate dimension. The regression matrix $\Basfun^j$ can then be expressed as a Kronecker product ${\Basfun^j=\Basfun^j_1\otimes\cdots\otimes\Basfun^j_P}$. For instance, the Gaussian \abbrRBF is an example of a product basis function and is separable across dimensions. In the \abbrGP description, this would correspond to using a product kernel, \emph{i.e.}, 
\begin{equation}
\kappa(\state,\state') = \prod_{l=1}^{P}\kappa([x]_p,[x]_p').
\end{equation}
If a $P$-dimensional square grid is assumed (\emph{i.e.}, $\Bfcenter_p$ is of size $m,~ \forall p$), using \cref{eq:bf:productbf} can reduce the number of basis function evaluations from $m^P$ to $Pm$. Unfortunately, factorizing compact {\abbrRBF}s in this way is usually difficult, which is easy to see from \cref{eq:basisfunction:crbf}. It can, however, be approximately factorized in this way for finding relevant basis functions.

The discussion is now, without loss of generality, temporarily restricted to two dimensions. Essentially, the {\abbrRBF}s considered here are circles in 2D, \emph{i.e.}, if both dimensions are considered simultaneously, the basis functions will yield a circle of non-zero values. If, instead, each dimension is evaluated separately, the basis functions will yield a square of non-zero values. As a square of side-length $a$ will always enclose a circle of diameter $a$, the necessary basis functions can be approximately computed using \cref{eq:bf:productbf}. See~\cref{fig:bfselection} for an illustration of the suggested basis function selection method. The suggested method for finding active basis functions and computing their values, henceforth referred to as fast \abbrCSRBF, clearly reduces the amount of necessary basis function evaluations. 
The drawback is a slight increase in memory requirement as the correlations between these weights need to be stored, as well as a restriction to square grids. It also increases $\tilde{n}_\weight^j$ by the set difference of the \abbrCSRBF and fast \abbrCSRBF basis function selection methods. However, as this set difference is small compared to the total number of basis functions, a performance increase is still expected. 
\begin{figure}[tb]
	\centering
	\includegraphics[width=\columnwidth]{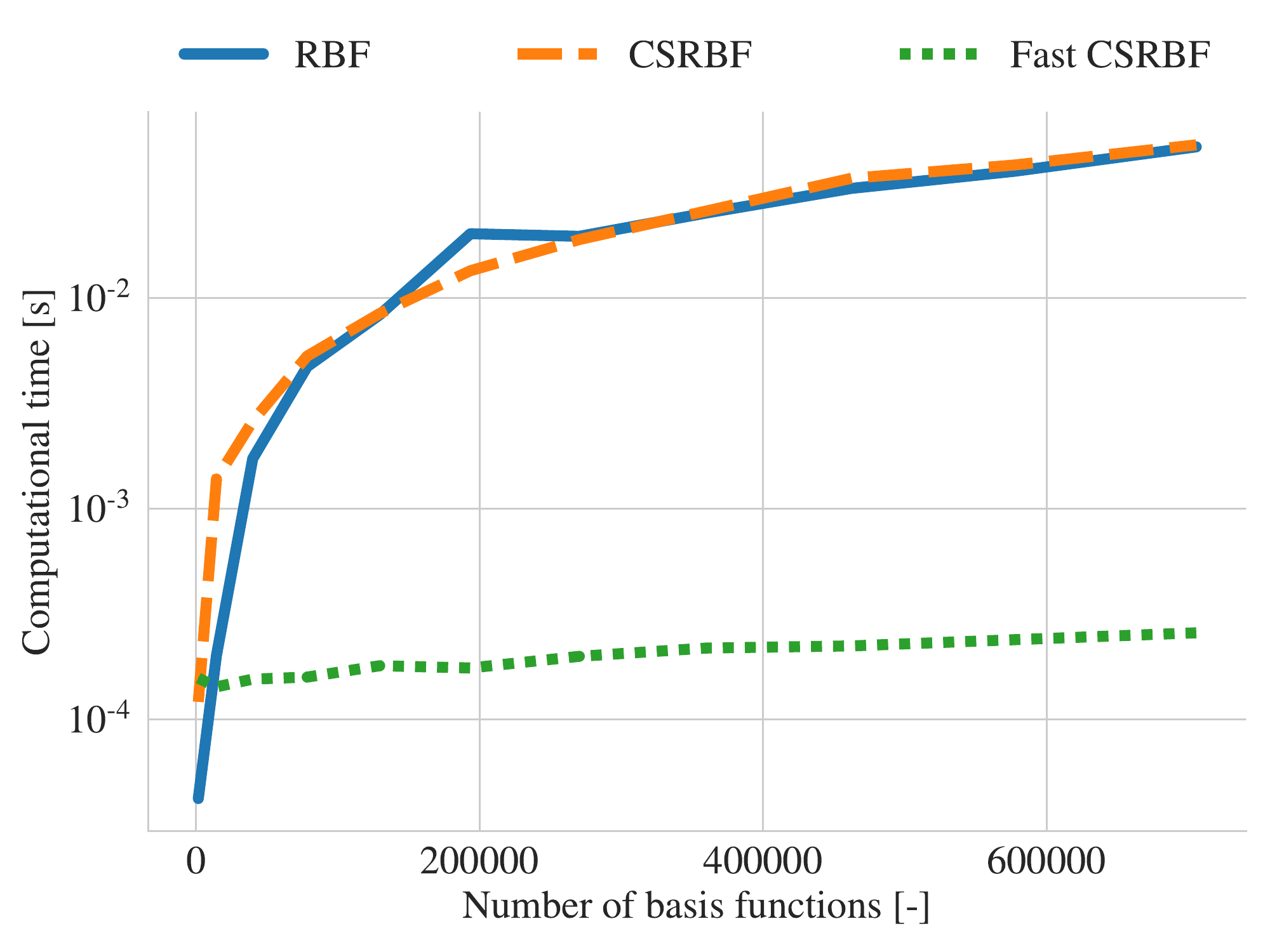}
	\caption{Evaluation time of $\bunf_k$ for increasing number of basis functions. Using the standard compact formulation sees no gain in evaluation time compared to the full \abbrRBF formulation since it is still necessary to evaluate all basis functions to determine which ones to use. The approximate compact method vastly improves over both other methods.}
	\label{fig:inputperf}
\end{figure}

To illustrate the reduction in computational time when evaluating $\bunf_k$ using the fast \abbrCSRBF method, a 2D basis function grid was constructed, \emph{i.e.}, $P=2$. A constant velocity model was then used together with a two-dimensional $\bunf_k$, \emph{i.e.}, $J=2$. The grid density was then gradually increased to increase the number of basis functions. The three approaches (\emph{i.e.}, full \abbrRBF, \abbrCSRBF and fast \abbrCSRBF), were then used to compute $\bunf_k$ in a single point $\state_k$. The different evaluation times are shown in~\cref{fig:inputperf}. It is clear that the fast \abbrCSRBF is superior, especially for large $n_\weight$. The evaluation times of the \abbrCSRBF and full \abbrRBF descriptions are approximately the same, which is not surprising given that both of the methods are evaluating all of the basis functions. 

\begin{figure}[t!]
	\centering
	\includegraphics[width=\columnwidth]{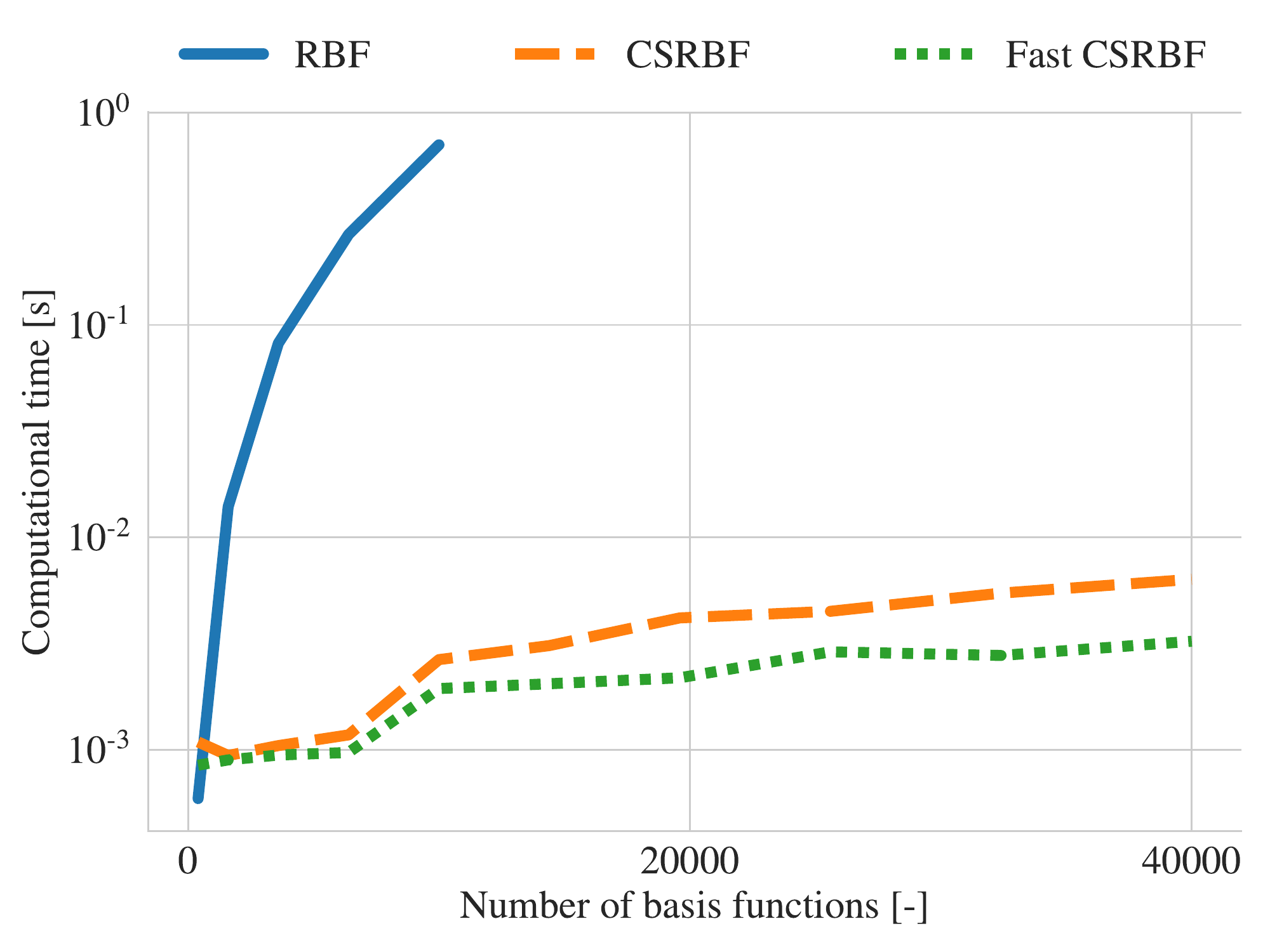}	
	\caption{Computational time as a function of the number of basis functions when performing one-step ahead prediction with a two-dimensional basis function grid and a constant velocity dynamical model augmented with a two-dimensional $\bunf_k$.} 
	\label{fig:propagateperf}
\end{figure}

The one-step-ahead prediction times for the complete model, \emph{i.e.}, predicting the state and covariance at the next time step were also compared for the three methods, see~\cref{fig:propagateperf}. After ${\sum_j n^j_\weight\approx10000}$, the full \abbrRBF caused memory overflow and was left out for further increases. The \abbrCSRBF and fast \abbrCSRBF descriptions have similar performance. However, the fast \abbrCSRBF description still has a clear advantage, because of the computationally efficient selection of active basis functions highlighted in~\cref{fig:inputperf}. Of course, the fast \abbrCSRBF description is limited to regular grids. However, using a regular grid alleviates the need for precise prior knowledge of the geometry of the environment and is instead limited to knowledge about the extent of the scenario. Hence, regular grids are potentially more practically useful than non-regular grids. Furthermore, both the fast \abbrCSRBF method as well as the ordinary \abbrCSRBF methods are nearly invariant to the total number of basis functions, see \cref{fig:propagateperf}. Hence, it is mainly the amount of available memory that limits the number of basis functions to be used.

\subsection{Choice of Basis Function Sets}\label{subsec:choiceofbfsets}
The set of basis function centers $\Bfcenter$ can be chosen rather arbitrarily. Let $\state_k\in\setX,~\forall k$ denote the input to $\basfun^j_i$. Then, the $\bfcenter_i^j$ are typically chosen from $\setX$, \emph{i.e.}, $\Bfcenter\subseteq\setX$. As the method described herein is concerned with basis functions with compact support, it is important that the basis function grid covers the entire domain of $\bunf_k$, as the function will be identically equal to $0$ outside the support of the basis functions.

Further, if $\bunf_k$ is multi-dimensional, then several basis function expansions are necessary. If the effects being modeled are similar in nature, such as the two-dimensional globally aligned acceleration of a vehicle, it is reasonable to use the same basis functions for all of the expansions (\emph{i.e.} $\bbasfun_k^1=\cdots=\bbasfun_k^J$). Since {\abbrRBF}s only depend on $\lVert\state_k-\bfcenter^j_i\rVert$, the same basis function values can be used for each separate dimension $j$ of $\bunf_k$. Hence, the number of basis function evaluations are reduced from $n_\weight$ to $n^j_\weight$. 

\subsection{Basis Function Weight Ordering}\label{subsec:weightordering}
Assuming $\bbasfun_k^1=\cdots=\bbasfun_k^J=\bbasfun_k$ (\emph{i.e.}, the same set of basis functions are used for all dimensions of $\bunf_k$), then two distinct options of weight ordering (\emph{i.e.}, the ordering of $\bweight_k$) exist. The first option is to stack all of the weights as $\bweight_k=\begin{pmatrix}(\bweight_k^1)^\top&\cdots&(\bweight_k^J)^\top\end{pmatrix}^\top$. This leads to
\begin{equation}
\label{eq:perf:stackedweights}
\bunf_k = (I\otimes\bbasfun^\top_k)\bweight_k.
\end{equation}
The second option is to stagger the weights (\emph{i.e.}, keep all weights $\weight_{k,i}^j$ associated with $\basfun_i$ together in $\bweight_k$). Then,
\begin{equation}
\label{eq:perf:staggeredweights}
\bunf_k = (\bbasfun^\top_k\otimes I)\bweight_k.
\end{equation}
Mathematically, the representations in \cref{eq:perf:stackedweights} and \cref{eq:perf:staggeredweights} are equivalent. However, depending on how the memory access is handled by the used programming language and hardware, these two representations result in significantly different execution times. For example, the representation in \cref{eq:perf:staggeredweights} results in a $3\times$ faster function evaluation in Python.

\subsection{Memory Usage}\label{subsec:memoryusage}
Thus far, we have mainly analyzed aspects relating to the computational complexity. However, the necessary storage also needs to be considered. Here, we will only focus on the memory S needed for the representation of $\bunf_k$ and leave out the memory requirements relating to $\bknf_k$, which are application-specific. Assuming that all numbers are stored with $d$ bits and an equally spaced grid is used so that $n^1_\weight=\cdots=n^J_\weight$, the mean vector $\hat{\bweight}$ will require $d\cdot n^j_\weight \cdot J$ bits, and if the symmetry is ignored, the covariance matrix $\Pw$ will require
\begin{equation}
\label{eq:storagecost}
S = d\cdot (n_\weight^j\cdot J)^2
\end{equation}
bits. Unfortunately, if the dynamical systems affected by $\bunf_k$ are allowed to travel over the entire support of $\bunf_k$, the matrices $\Pw,~\Pxw$ and $\Pwx$ will be dense and hence, the storage required is fixed. Thus, the compact basis functions offer no gains in terms of storage complexity.

\section{Complete Model and Estimation}\label{sec:modeling}

To summarize the suggested complexity reduction, explicit design choices must be made. Assume that the same set of basis functions are used for each dimension of $\bunf_k$, \emph{i.e.}, $\bbasfun^1_k=\cdots=\bbasfun^J_k$. This implies that the indicator matrices $\Ind^1_k=\cdots=\Ind^J_k=\Ind_k$. Also, assume that the weights are stacked according to \cref{eq:perf:stackedweights}. The function $\bunf_k$ is then described by
\begin{equation}
\bunf_k=\Basfun_k\bweight_k,
\end{equation}
where $\Basfun_k=\left(\vec{I}\otimes\bbasfun_k^\top\Ind_k^\top\Ind_k\right)$.
There are two remaining aspects necessary to consider. First, the dependency of $\Basfun_k$ on $\state_k$ needs to be specified. Here, consider a general transformation of $\state_k$, \emph{i.e.}, 
\begin{equation}
\bunf_k = \bunf_k(\indi(\state_k),\inp_k),
\end{equation}
where $\indi$ is some transformation of $\state_k$, for instance, ${\indi(\state_k)=\vec{D}_k\state_k}$ if the dependency is linear. The transformation $\indi$ is assumed known and is specified a priori. The final model is then given by
\begin{subequations}
\label{eq:mod:fullmodel}
\begin{align}
\label{eq:model:state}
\state^e_{k+1} &\triangleq 
\bknf^{e}_k(\state^{e}_k,\inp_k,\vec{w}^{e}_k, \bunf_k)\\
\bknf^{e}_k &\triangleq
\begin{pmatrix}
\bknf_k\left(\state_k, \inp_k, \vec{w}_k, \bunf_k\right)\\
\bweight_k + \bnoise_k
\end{pmatrix}\\
\bunf_k &= \left(\vec{I}\otimes\bbasfun_k^\top(\indi(\state_k))\Ind_k^\top\right)\Ind_k\bweight_k\triangleq\Basfun_k\bweight_k\label{eq:model:unknownfunc}\\
\label{eq:model:observation}
\obs_k &= \vec{h}_k(\state_k,  \inp_k, \vec{e}_k),
\end{align}
\end{subequations}
where $\vec{e}_k,~\vec{w}_k$ and $\bnoise_k$ are mutually independent white noise processes with covariance matrices $\vec{R}_k,~\vec{Q}_k,$ and $\bm{\Sigma}_k$, respectively.

\subsection{Filter Modifications}\label{subsec:filtermodifications}
The \abbrEKF described in \cref{subsec:problemformulation:compcomplexity} can be used to estimate $\state_k$ and $\bweight_k$ for every $k$. 
However, to fully exploit the new model structure and the computational complexity reductions it enables, the filter equations must be altered. Starting with the time update, note that from \cref{eq:model:state}, the mean update essentially remains the same apart from $\bunf_k$ now being defined by \cref{eq:model:unknownfunc}. The covariance update \cref{eq:ekf:timeupdate:covariance} is altered according to
\begin{subequations}
\begin{align}
\Px\pkk={}&\Fx\Px\kk{\Fx}^\top+\Fx\Pxw\kk\Ind_k^\top{\Fw}^\top\nonumber\\
&{+}\:\Fw\Ind_k\Pwx\kk{\Fx}^\top+\Fw\Ind_k\Pw\kk\Ind_k^\top{\Fw}^\top+\vec{Q}_k\\
\Pxw\pkk={}&\Fx\Pxw\kk+\Fw\Ind_k\Pw\kk.
\end{align}
\end{subequations}
Hence, the bulk of the computational complexity relating to the product $\Fw\Pw\Fw^*$ has been reduced to $\bigO(\tilde{n}^2_\theta)\ll\bigO(n^2_\theta)$. The measurement update alterations are a bit more involved as some approximations are involved. First, we must begin by noting that $\Lw=\Pxw\kkm{\Hx}^\top\vec{S}_k^{-1}$, and as such, the computational complexity can not be reduced while remaining with an exact algorithm as $\Pxw\kkm$ is dense after some time (given that the system has been influenced by all basis functions). Hence, approximations are needed to reduce the computational complexity. Here, we draw inspiration from Simultaneous Localization and Mapping, and more precisely \cite{Julier2001}, where a sparse weight Kalman filter is derived. Particularly, a gain matrix $\vec{K}_k$, which updates but a few of the states, is derived and shown to minimize the trace of the resulting covariance matrix. Here, the measurement update is restricted to recently active basis functions, \emph{i.e.},
\begin{equation}
\vec{K} = \begin{bmatrix}
{\Lx}^\top & (\Ind_k^\top\Ind_k\Lw)^\top
\end{bmatrix}^\top.
\end{equation}
Hence, $\vec{K}_k$ is sparse and all that remains is to alter the measurement equations. Essentially, all that is required is to replace $\Lw$ with $\Ind_k^\top\Ind_k\Lw$ which will leave only the elements of $\Lw$ that are non-zero in $\Ind_k$. To summarize, the new measurement update of the \abbrEKF becomes
\begin{subequations}
	\begin{align}
	\hat{\bweight}\kk ={}& \hat{\bweight}\kkm+\Ind_k^\top\Ind_k\Lw(\obs_k-\Hx\hat{\state}\kkm)\\
	\Pxw\kk ={}& (\vec{I}-\Lx \Hx)\Px\kkm(-\Ind_k^\top\Ind_k\Lw\Hx)^\top\nonumber\\
	&{+}\:(\vec{I}-\Lx\Hx)\Pxw\kkm+\Lx\vec{R}_k{\Lw}^\top\Ind_k^\top\Ind_k\\
	\Pw\kk ={}& \Pw\kkm + \Ind_k^\top\Ind_k\Lw\Hx\Px\kkm{\Hx}^\top{\Lw}^\top\Ind_k\Ind_k^\top\nonumber\\
	&{-}\:\Ind_k^\top\Ind_k\Lw\Hx\Pxw-\Pwx{\Hx}^\top{\Lw}^\top\Ind_k\Ind_k^\top\nonumber\\
	&{+}\:\Ind_k^\top\Ind_k\Lw\vec{R}_k{\Lw}^\top\Ind_k\Ind_k^\top.
	\end{align}
\end{subequations}
Hence, similar to the time-update, the computational complexity relating to the product $\Lw\vec{R}_k{\Lw}^\top$ has been reduced to $\bigO(\tilde{n}^2_\theta)\ll\bigO(n^2_\theta)$. Note that the reduction in computational complexity hinges on how $\Ind_k$ is used. In a practical implementation, the direct use of $\Ind_k$ \textit{increases} the complexity; therefore, the user must instead devise a way of efficiently indexing into $\bbasfun_k$ and $\bweight_k$ and thus indirectly use $\Ind_k$. However, this has been left out as this depends on the choice of implementation language.

Lastly, to quantify the number of basis functions used in each iteration, assume that the basis function grid is regular and of dimension $P$ with grid density $\delta_c$ in all dimensions. Further, assume that the support of the basis functions is $\alpha$, and that the current state $\state_k$ is equal to one of the basis function centers. Then, the number of active basis functions with the fast \abbrCSRBF selection method is found by noting that the active region is an $n$-dimensional cube with side length $\alpha$. The upper bound of active number of basis functions is thus
\begin{equation}\label{eq:nactivebasis}
\tilde n_\weight \leq \left(\frac{2\alpha}{\delta_c}+1\right)^P,
\end{equation}
which is useful for determining a reasonable tuning of $\alpha$ and $\delta_c$, depending on the available computational resources. Note that if $\state_k$ is not equal to one of the basis function centers, or if $\state_k$ is close to the boundary of the basis function grid, then the inequality is strict.

\subsection{Parameter Selection}
The parameter selection in the model is a bit nuanced as it has two distinct effects: (i) it influences the state error, and (ii) it affects what types of functions can be represented as well as the convergence rate of the basis function expansion. The parameters to choose are the prior motion model, the domain of $\bunf$, the basis function grid density $\delta_c$ and support $\alpha$, the noise covariances $\vec{Q}_k,~\vec{R}_k$, and $\bm{\Sigma}_k$, and the priors $\hat{\state}_0,~\Px_0,~\Pxw_0,~\bweight_0$, and $\Pw_0$. The parameters are problem dependent and it is hard to give general advice for the selection. Most of the parameters, excluding the prior motion model and the domain of $\bunf$, can be found through, \emph{e.g.}, maximum likelihood methods applied to historical data, or they can be manually selected based on expert knowledge. An intuition about the parameters is given here, and we later illustrate the selection process in connection to the numerical examples.

The prior motion model should generally be chosen to, as accurately as possible, represent the system at hand, so that the basis function expansion does not need to represent already known system properties. The basis function support $\alpha$ and grid density $\delta_c$ affect the computational complexity of the algorithm, but also what types of functions that can be represented. The support $\alpha$ completely determines the range of the spatial correlations that can be captured by the model. The density $\delta_c$ determines how quickly the learned function can vary with the input. The noise covariances $\vec{Q}_k,~\vec{R}_k$, and $\bm\Sigma_k$ affect both the tracking error and the learning rate of the basis function weights. Further, $\bm\Sigma_k$ also determines how fast the model ``forgets'' past information. Lastly, the prior $\Pw_0$ acts as a regularizer which keeps the basis function weights close to the prior and is key to keeping the learning rate slow.

Some of the parameters can also possibly be learned online, for instance through further random walk assumptions. Particularly interesting, as alluded to in \cref{subsec:compactsupport}, is the \abbrBF support $\alpha$ and the grid density $\delta_c$. These two parameters completely determine the computational complexity of the method. Hence, if these are to be learned online, it will be important to impose restrictions on the values they can take, to remain with a real-time applicable algorithm. Another possibility is to adapt the noise covariances $\vec{Q}_k$ and $\bm\Sigma_k$ online, as in, \emph{e.g.}, \cite{Bavdekar2011:EMEKF,Goodwin:2005:EMStateEstimation}, where an expectation maximization approach was taken to estimate the noise covariances over time.

\section{Numerical Examples}\label{sec:experiments}

To illustrate the application of the proposed joint state inference and model learning approach, we will present three simulation examples. The first example highlights a key property of the presented method, namely the effect of including a prior motion model. The second example estimates the longitudinal tire-friction of a car and is designed to quantitatively compare the estimation performance of a standard \abbrEKF using the {\abbrSSM}+\abbrRBF model and the sparsely weighted \abbrEKF with the {\abbrSSM}+\abbrCSRBF model. The last example estimates the accelerations of vehicles throughout a three-way intersection and is designed to illustrate the reduction in computational complexity when using the fast \abbrCSRBF description compared to the standard \abbrRBF description of $\bunf_k$ in a larger setting. In the two latter examples, the method is compared to an \abbrEKF with the {\abbrSSM}+\abbrRBF model which in all essence is identical to the {\abbrSSM}+\abbrGP model used in \cite{Kullberg2020}.

\subsection{Influence of Prior Motion Model}\label{subsec:experiments:constantvel}
One particular property of the model \cref{eq:int:filtproblem} is that it inherits an initial estimation performance, in a \textit{root mean squared error} (\abbrRMSE) sense, from the chosen prior motion model. Further, the model can improve upon that performance in cases where the prior motion model does not accurately describe the system. These properties are highlighted in the following example.

\subsubsection{Modeling}
Three different models, $\bknf_k^{(a)},~\bknf_k^{(b)}$, and $\bknf_k^{(c)}$ are considered, which are of the following form
\begin{subequations}
	\begin{align}
	\label{eq:bknfk1}
	\state^{(a)}_{k+1} &= \begin{bmatrix}
	1 & 1\\0 & 1
	\end{bmatrix}\state^{(a)}_k+\begin{bmatrix}
	1/2\\1
	\end{bmatrix}\vec{w}^{(a)}_k\\
	\label{eq:bknfk2}
	\state^{(b)}_{k+1} &= \begin{bmatrix}
	1 & 1\\0 & 1
	\end{bmatrix}\state^{(b)}_k+\begin{bmatrix}
	1/2\\1
	\end{bmatrix}(\bunf^{(b)}_k({\state_k^{(b)}})+\vec{w}^{(b)}_k)\\
	\label{eq:bknfk3}
	\state^{(c)}_{k+1}&=\bunf^{(c)}_k(\state_k^{(c)})+\vec{w}^{(c)}_k.
	\end{align}
\end{subequations}
Also, assume that the first state component (the position) is measurable, \emph{i.e.}, the sensor model is
\begin{equation}
y = [x_k^{(i)}]_1+e^{(i)}_k,\quad i\in[a,~b,~c].
\end{equation}
Thus, the models \cref{eq:bknfk1,eq:bknfk2,eq:bknfk3} are a \abbrCV model, a \abbrCV model with basis function expansion augmentation, and a basis function expansion description, respectively. 
\subsubsection{Parameter Selection}
The three models \cref{eq:bknfk1,eq:bknfk2,eq:bknfk3} are tuned identically. The domain of $\bunf$ is chosen to cover the state-space of the simulations. The grid density $\delta_c=\SI{1}{\meter}$ and the basis function support is chosen as $\alpha=10$ to have at least a few basis functions active in each time step. The measurement and process noise have variances $\vec{Q}=0.01$ and $\vec{R}=0.01$, respectively. The priors $\Pw_0=0.1\vec{I}$ and $\Px_0=\vec{I}$.

\subsubsection{Results}
To illustrate the benefits of including the prior motion model in \cref{eq:bknfk2}, two different scenarios are considered. The objective is to estimate the position of the system with minimal error and the performance is compared through the positional \abbrRMSE. A Kalman filter is applied to \cref{eq:bknfk1} and an \abbrEKF to \cref{eq:bknfk2,eq:bknfk3}, respectively.
The two considered scenarios are:
\begin{enumerate}
	\item The ``true'' system is assumed to be given by \cref{eq:bknfk1}. It is simulated 100 time steps with a process noise with variance $\vec{Q}=0.01$, and the measurement error has variance $\vec{R}=0.01$.
	\item The ``true'' system is assumed to be given by a model similar to \cref{eq:bknfk2}, where the ``true'' $\bunf_k^{(2)}$ is assumed to be given by $\frac{1}{2}\sin(\frac{1}{25}\pi [\state_k]_0)+0.01$. This choice is to keep the states somewhat close to the origin so that the number of basis functions necessary to describe the system using \cref{eq:bknfk3} is kept low.
\end{enumerate}
The simulations were run 50 times and the mean positional \abbrRMSE is reported in \cref{tab:example1rmse}. In the first scenario, the models \cref{eq:bknfk1,eq:bknfk2} have essentially identical estimation performance whereas \cref{eq:bknfk3} has a higher \abbrRMSE. This indicates that the prior motion model included in \cref{eq:bknfk2} provides an initial estimation performance, as opposed to \cref{eq:bknfk3}. For the second scenario, model \cref{eq:bknfk3} performs worst. The model \cref{eq:bknfk2} performs better than both \cref{eq:bknfk1} and \cref{eq:bknfk3}, which again indicates an initial estimation performance, but also that the model improves over time, as opposed to \cref{eq:bknfk1}. Clearly, the prior motion model provides a minimal estimation performance, in an \abbrRMSE sense, while the basis function expansion augmentation provides a means of improving over time.
Note that this does not mean that including a prior motion model will always perform better than just using a basis function expansion. Depending on how well the assumptions of the model on the function $\bunf_k$ fit the actual scenario, one model or the other may perform better.
\begin{table}[t!]
	\centering
	\caption{Mean \abbrRMSE for \cref{subsec:experiments:constantvel}\label{tab:example1rmse}}
	{\renewcommand{\arraystretch}{1.3}%
		\begin{tabular}{P{.15\columnwidth}P{.2\columnwidth}P{.2\columnwidth}}\toprule
			\textbf{Model} & \multicolumn{2}{c}{\textbf{Mean \abbrRMSE}}\\
			& \textbf{Scenario 1)} & \textbf{Scenario 2)}\\\midrule
			\cref{eq:bknfk1} & $0.09$ & $0.18$\\
			\cref{eq:bknfk2} & $0.09$ & $0.09$\\
			\cref{eq:bknfk3} & $4.14$ & $12.69$
			\\\bottomrule
	\end{tabular}}
\end{table}

\subsection{Longitudinal Tire-Friction Estimation}
The second example deals with longitudinal tire-friction estimation in a two-wheel drive vehicle. The scenario is chosen to illustrate that the approximate algorithm presented here is competitive, in a function estimation \abbrRMSE sense, compared to using an exact algorithm, see \cite{Kullberg2020}.

\subsubsection{Modeling}
\begin{table}[!t]
	\centering
	{\renewcommand{\arraystretch}{1.3}%
		\caption{Longitudinal friction model/simulation parameters\label{tab:tirefriction:parameters}}
		\begin{tabular}{P{.1\columnwidth}P{.2\columnwidth}m{.4\columnwidth}}\toprule
			\textbf{Parameter} & \textbf{Value} & \textbf{Description}\\\midrule
			$\delta_c$ & $0.025$ & Grid spacing\\
			$\alpha$ & $0.15$ & \abbrCSRBF support\\
			$l$ & $0.01$ & \abbrRBF length scale\\
			$\vec{R}$ & \diag[0.1,~0.01] & Measurement noise covariance\\
			$\hat{x}_0$ & 0 & Prior state estimate\\
			$\Px_0$ & $10^{-6}$ & Prior state error covariance\\
			$\Pw_0$ & $10^{-5}\vec{I}$ & Prior weight error covariance\\
			$q$ & $1$ & Process noise variance\\
			$\bm\Sigma$ & $10^{-8}\vec{I}$ & Weight noise covariance\\
			$T_s$ & $\SI{40}{\milli\second}$ & Sampling interval\\
			$l_r$ & $\SI{1.6}{\meter}$ & Length from \abbrCG to rear axle\\
			$l_f$ & $\SI{1.4}{\meter}$ & Length from \abbrCG to front axle\\
			$m$ & $\SI{1000}{\kilogram}$ & Vehicle mass\\
			$g_0$ & $\SI{9.81}{\meter\per\second^2}$ & Nominal gravity\\
			$B$ & $11.7$ & Pacejka stiffness parameter\\
			$C$ & $1.69$ & Pacejka shape parameter\\
			$D$ & $1.2$ & Pacejka peak parameter\\
			$E$ & $0.377$ & Pacejka curvature parameter
			\\\bottomrule
		\end{tabular}
	}
\end{table}
\begin{figure*}[!t]
	\centering
	\includegraphics[width=.6\textwidth]{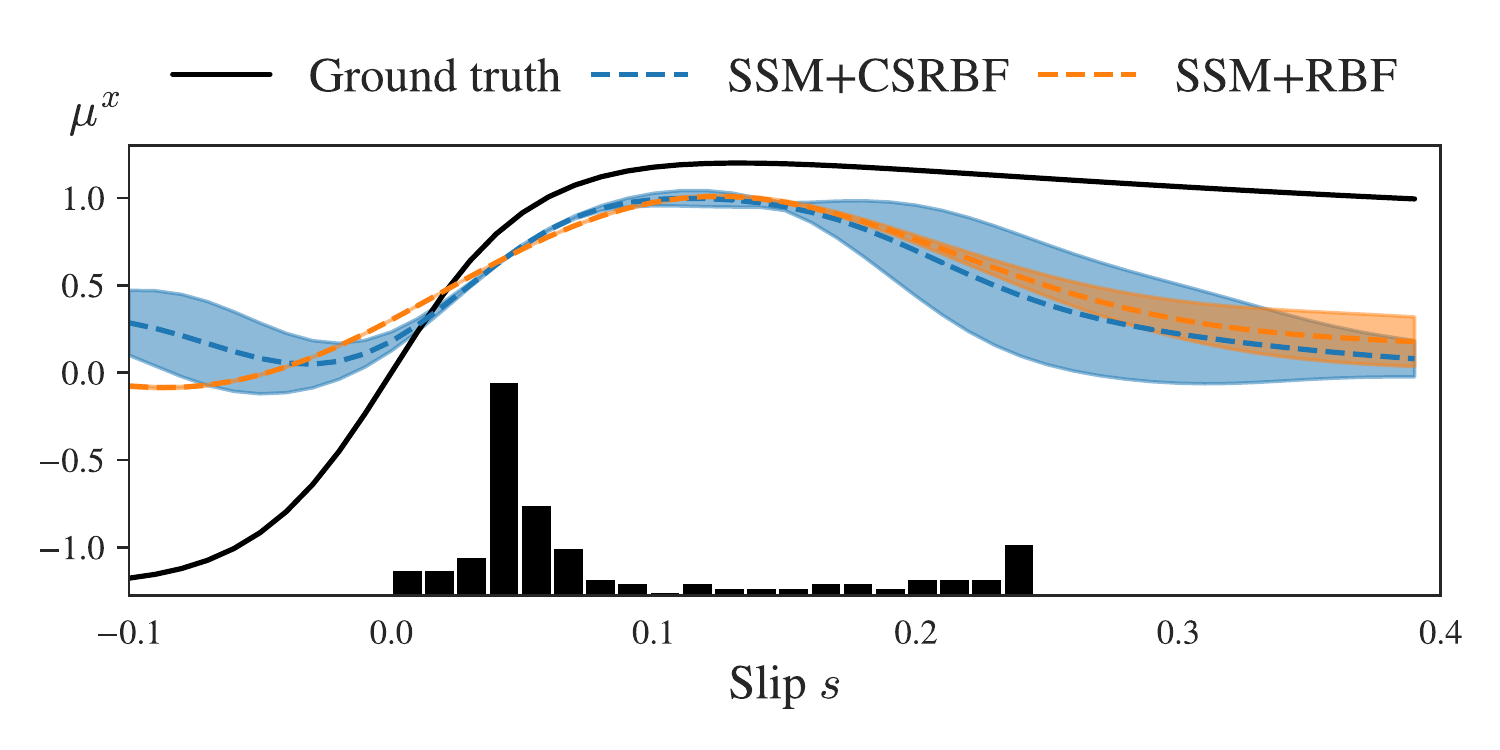}
	\caption{Longitudinal friction force estimate with the proposed method and the {\abbrSSM}+\abbrRBF method described in \cite{Kullberg2020}. The frequency of the true slip is plotted as a black histogram, displaying in what regions the methods ``should'' have data available. The true slip is not known but estimated online. The confidence intervals is a $3\sigma$ interval constructed from the $50$ \abbrMC realizations.\label{fig:lonfrictionestimate}}
\end{figure*}
The motion model is chosen as the commonly used single-track model, see \emph{e.g.} \cite{Rajamani2011a} for details. A detailed derivation of the motion model can be found in the supplementary material and it is summarized here. The following assumptions are made
\begin{itemize}
	\item The steering angle is small, such that the lateral and angular velocities are negligible.
	\item The pitch dynamics is negligible, so that the vertical load on the wheels is constant.
	\item The angular velocity of the wheels is assumed measurable.
	\item The vehicle is assumed to be front-wheel drive, so the longitudinal velocity is measurable through the free-rolling rear wheels.
	\item The vehicle is equipped with an accelerometer capable of measuring the longitudinal acceleration, a commonly occurring automotive sensor \cite{Berntorp2019a}.
\end{itemize}
The tire friction is modeled as a basis function expansion and the complete, discretized, model is thus given by
\begin{subequations}
	\label{eq:lonfrictionmodel}
	\begin{align}
	\label{eq:lonfrictionmodel:dynamics}
	x_{k+1} &= x_k+T_s\underbrace{\frac{g_0l_f}{l_r+l_f}}_{G}\unf_{f,k}+w_k\\
	\unf_{f,k} &= \bbasfun^\top(s_k(x_k))\Ind_k^\top\Ind_k\bweight_k\\
	\label{eq:lonfrictionmodel:wheelslip}
	s_k(x_k) &=\frac{r_w\omega_{f,k}-x_k}{x_k}\\
	\bweight_{k+1} &= \bweight_k+\bnoise_k\\
	\label{eq:lonfrictionmodel:sensor}
	\obs_k &= \begin{bmatrix}
	G\unf_{f,k}\\ x_k
	\end{bmatrix}+\vec{e}_k.
	\end{align}
\end{subequations}
Here, $x_k$ is the longitudinal velocity of the vehicle, $g_0$ is the nominal gravity, and $l_r$ and $l_f$ are the distances from the \emph{center of gravity} (\abbrCG) to the rear and front axle, respectively. Further, $r_w$ and $\omega_{f,k}$ are the radius and angular velocity of the front wheels, respectively. The wheel slip is denoted $s_k$ and $T_s$ is the sampling interval, which is here $T_s=\SI{40}{\milli\second}$. Furthermore, $w_k,~\bnoise_k$ and $\vec{e}_k$ are mutually independent white noise processes with covariances $q,~\bm\Sigma$ and $\vec{R}$, respectively. Lastly, the static vehicle parameters are given in \cref{tab:tirefriction:parameters}.

\subsubsection{Parameter Selection}
As the motion model has been fixed, the basis function grid can be chosen. The interest here lies in estimating the main characteristics of the tire-friction curve and the domain of $\bunf_k$ is thus chosen as $s_k\in[-0.5,~0.5]$. The noise covariance $\vec{R}_k=\vec{R}$ is chosen according to the simulated noise levels in the sensors. The prior $\hat{x}_0=0$ and $\Px_0=10^{-6}$ as it is known that the vehicle starts from a standstill. Since the unknown function $\unf_{f,k}$ influences both the motion model and the sensor model, the prior $\Pw_0$ is chosen to be small, to keep the model from adapting too quickly, which can otherwise lead to divergence. To not place too strict assumptions on the smoothness of the tire friction curve, the grid density $\delta_c=0.025$ and basis function support $\alpha=0.15$, which keeps the measurements from influencing function estimates far from the current slip value. The \abbrRBF length scale is chosen as $l=0.01$, as this proved to yield the best performance in that case. The covariances $q$ and $\bm\Sigma$ are finally tuned to achieve a low \abbrRMSE w.r.t. to the function estimate. The two methods are tuned independently for the lowest possible \abbrRMSE and hence, the parameters differ slightly. The complete set of parameters for the {\abbrSSM}+\abbrCSRBF model is presented in \cref{tab:tirefriction:parameters}.

\subsubsection{Results}
The scenario was simulated in MATLAB using the Simscape Driveline testbed with the true tire friction given by the Pacejka tire model\cite{Pacejka}
\begin{equation}
\mu=D\sin(C\arctan(Bs-E(Bs-\arctan(Bs)))),
\end{equation}
where the parameters $B,C,D$, and $E$ are chosen to correspond to an asphalt surface, see \cref{tab:tirefriction:parameters}. The scenario consists of five accelerations from standstill to a longitudinal velocity of $v\approx \SI{20}{\meter\per\second^{}}$ to excite the system into the nonlinear region of the slip curve, since it is approximately linear for small slip values. 100 accelerations from standstill were simulated and Gaussian noise was added to the longitudinal acceleration and wheel angular velocity measurements, with standard deviation $0.1$ and $0.01$, respectively. Five of these accelerations are then sampled during evaluation to learn the function $\unf_{f,k}$ while jointly estimating the vehicle longitudinal velocity. This was done $50$ times, and the resulting function estimates are presented in \cref{fig:lonfrictionestimate}, where the confidence interval is a $3\sigma$ interval constructed from the $50$ realizations. The frequency of true slip is also visualized as histogram in the bottom of the figure. Note that the true slip is not known, but estimated online, and as such, the histogram is not a one-to-one mapping to the actual input to the basis function expansions. Neither of the methods recover the true friction force curve exactly, but both of them capture the general characteristics of the friction curve. The error was computed at the parts of the curve where there should be measurements, \emph{i.e.}, at the locations of the true slip. The mean \abbrRMSE for the two respective methods is presented in \cref{tab:longitudinalfriction:rmse}.
\begin{table}[!t]
	\centering
	\caption{\abbrRMSE of the estimated function}
	\label{tab:longitudinalfriction:rmse}
	\renewcommand{\arraystretch}{1.3}%
	\begin{tabular}{P{.25\columnwidth}P{.4\columnwidth}}\toprule
		\textbf{Model} & \textbf{RMSE} (std)\\\midrule
		{\abbrSSM}+\abbrCSRBF & $0.305~(\pm 0.0368)$\\
		{\abbrSSM}+\abbrRBF & $0.278~(\pm 0.0046)$ 
		\\\bottomrule
	\end{tabular}
\end{table}

\begin{figure}[!t]
	\centering
	\includegraphics[width=\columnwidth]{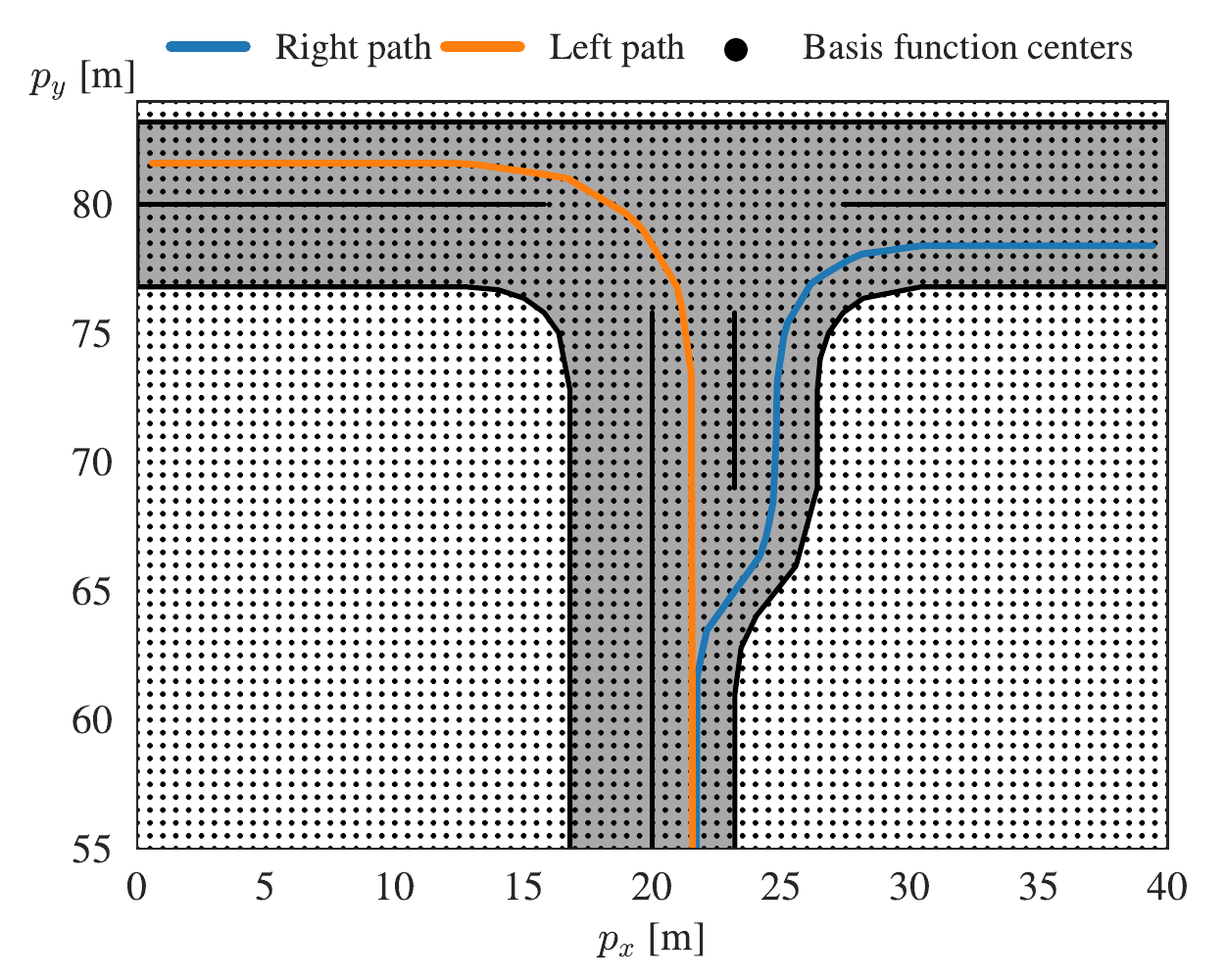}
	\caption{Three-way intersection simulation scenario. Cars approach on the middle road and turn left or right with equal probability. The two paths are visualized as well as the centers of the basis functions. }
	\label{fig:threewayintersection}
\end{figure}

\subsection{Three-Way Intersection Simulation}\label{subsec:threewayintersection}
The third example is similar to the one explored in \cite{Kullberg2020}. The scenario is a simulation of the three-way intersection shown in~\cref{fig:threewayintersection}. Cars approach from the bottom road, starting at $p_y=0$ (outside the image), and turn left or right with equal probability. In \cite{Kullberg2020}, we used a {\abbrSSM}+\abbrGP model parameterized by inducing points to model the dynamics of the cars. The inducing points were placed only over the actual road area to reduce computational burden and make the model viable in that specific scenario. Hence, there was a requirement on exact prior knowledge of the environment. As pointed out in \cref{subsec:findingbasisfunctions}, this requirement has been alleviated because the method described herein is \textit{nearly} invariant to the total number of basis functions. Hence, we use a simpler regular grid, and the only prior information needed is the total extent of the intersection.

\subsubsection{Modeling}
The prior motion model is chosen as a \abbrCV motion model, \emph{i.e.}, the complete motion model is
\begin{subequations}
	\label{eq:model:dynamicmodel}
	\begin{align}
	\state_{k+1} &= \vec{F}_k\state_{k}+\vec{G}_k\bunf_k+\vec{G}_k\vec{w}_k\\
	\vec{F}_k &= \begin{bmatrix}
	1 & T_s\\
	0 & 1
	\end{bmatrix}\otimes \vec{I} \qquad
	\vec{G}_k = \begin{bmatrix}
	{T_s}^2/2\\
	T_s
	\end{bmatrix}\otimes \vec{I},
	\end{align}
\end{subequations}
with $\bunf_k$ given by \cref{eq:model:unknownfunc}. Moreover, $T_s$ is the sampling interval for the system(s). Information about the input $\inp_k$ is assumed unknown; hence, $\inp_k=0,~\forall k$. Also, note that $\bunf_k$ is pre-multiplied by $\vec{G}_k$, which will render $\bunf_k$ to be interpretable as the vehicle acceleration throughout the intersection. Further, the necessary derivatives are given by
\begin{subequations}
	\begin{align}
	\Fx &= \vec{F}_k+\frac{\vec{G}_k\Basfun_k\hat\bweight\kk}{\partial \state\kk}\bigg\lvert_{\state\kk=\hat\state\kk}\\
	\Fw &= \vec{G}_k\Basfun_k(\hat\state\kk),
	\end{align}
\end{subequations}
where $\Basfun_k(\hat\state\kk)=\left(\vec{I}\otimes\bbasfun_k^\top(\indi(\hat\state\kk))\Ind_k^\top\Ind_k\right)$.

For the observation model, it is assumed that the position of the vehicle is measurable, albeit with some noise, \emph{i.e.},
\begin{subequations}
	\begin{align}
	\vec{h}_k &= \vec{H}_k\state_{k}+\vec{e}_k\\
	\vec{H}_k &= \begin{bmatrix}
	\vec{I} & \vec{0}
	\end{bmatrix}\qquad \vec{R}_k=\sigma_e^2 \vec{I},
	\end{align}
\end{subequations}
where $\sigma_e^2$ is the measurement noise variance. Note that a linear observation model implies $\nabla_{\state\kk}\vec{h}_k=\vec{H}_k$.

\subsubsection{Parameter selection}
The basis function grid spacing is selected $\delta_c=\SI{1}{\meter}$ primarily to allow the evaluations to run in reasonable time since the Gaussian {\abbrRBF}s otherwise cause a huge computational burden. Further, it could also be argued that the acceleration of a vehicle does not change rapidly enough to motivate a grid spacing less than $\delta_c=\SI{1}{\meter}$, thus justifying the choice. The support of the {\abbrCSRBF}s and length scale of the Gaussian {\abbrRBF}s is chosen as $\alpha=5$ and $l=1$, respectively. The length scale of the Gaussian {\abbrRBF}s is justified by the same argument as the grid spacing. The support of the {\abbrCSRBF}s is chosen as $\alpha=5$ to include at least a few basis functions in each time step. The measurement noise variance is set to $R=0.2$, which can be achieved through, \emph{e.g.}, differential \abbrGPS \cite{Skog2009}. The prior $\hat\state_0$ is set to the true initial position and velocity and $\Px_0=0.1\cdot\vec{I}$. In a real scenario, these can be found through the first two measurements and a simple numerical difference. The prior $\Pw_0=0.01\vec{I}$, which is significantly higher than the tire friction example, but is made possible by the inclusion of a prior motion model. The process noise covariance $\bm\Sigma_k=\vec{0}$, as the vehicles travel at approximately the same velocity, and the acceleration is thus not expected to change. The process noise covariance $\vec{Q}_k=\vec{Q}$ is then tuned to achieve a low \abbrRMSE w.r.t. the position and velocity estimates and was finally chosen as $\vec{Q}=0.1\cdot\vec{I}$. Furthermore, let the modeled acceleration depend only on the position of the vehicle, \emph{i.e.}, $$\indi(\state_k)=\begin{bmatrix}\vec{I} & \vec{0}\end{bmatrix}\state_k.$$
One might argue that normal acceleration also depends on the absolute velocity, but in this study, it is assumed that the vehicles travel at approximately the same speed throughout the intersection. Lastly, the sampling interval is set to $T_s=0.2$. All of the simulation parameters are summarized in \cref{tab:3way:modelparameters}.

\subsubsection{Results}
The intersection simulation was constructed and run in the SUMO microscopic traffic simulator \cite{SUMO2018}. A wide variety of vehicles were simulated and then sampled during estimation/learning, see \cref{tab:3way:sumoparameters} for a summary of the vehicle parameters. 150 vehicles were used for each of the 10 simulations. Further, as a tracking comparison, a common \abbrCV model is used with identical parameter tuning to both the {\abbrSSM}+\abbrCSRBF model as well as the {\abbrSSM}+\abbrRBF model.

\begin{figure*}[!t]
	\centering
	\includegraphics[width=1\textwidth]{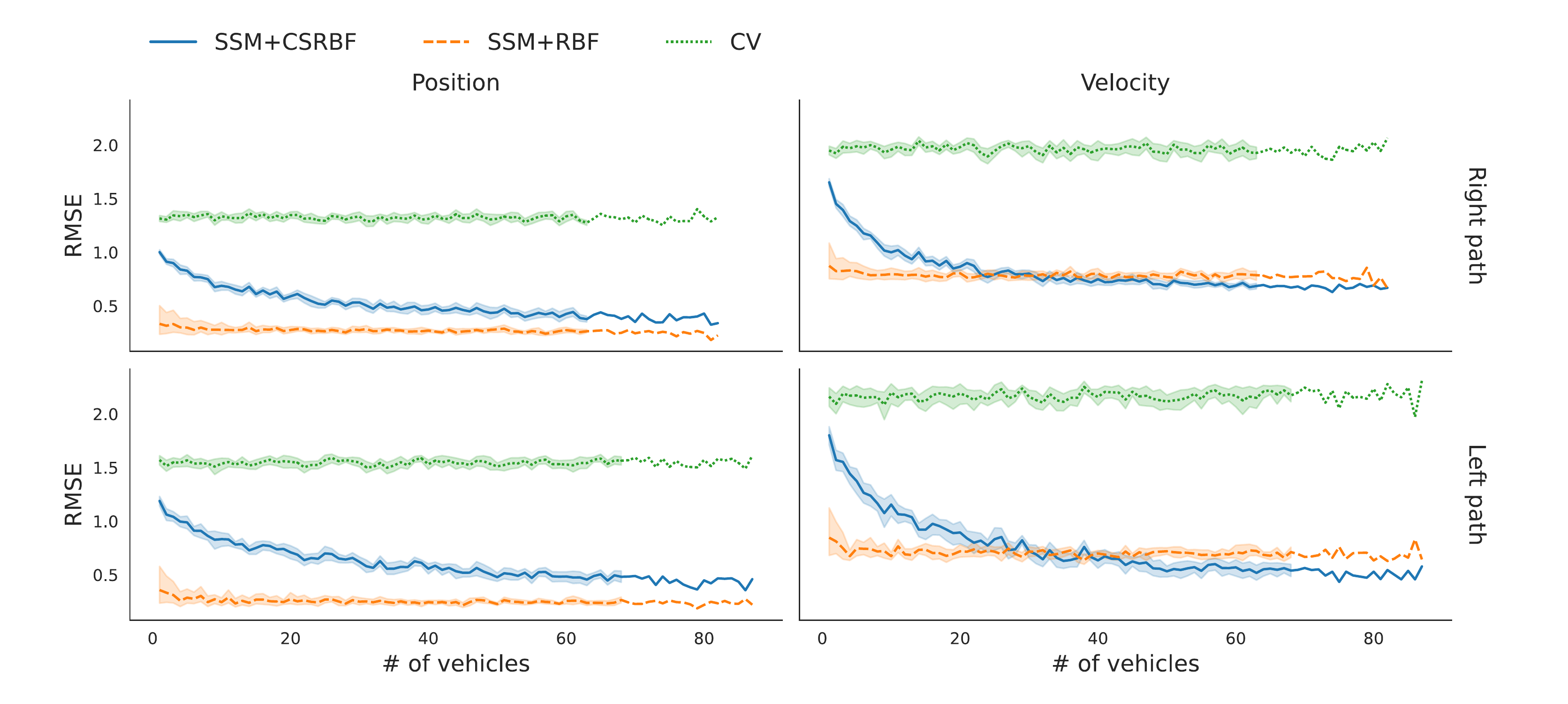}
	\caption{\label{fig:intersectionrmse}\abbrRMSE with an increasing number of vehicles, \emph{i.e.}, observations. The left column is the \abbrRMSE w.r.t. the position and the right column w.r.t. the velocity. The first row is w.r.t. the right path and the second row w.r.t. the left path. The {\abbrSSM}+\abbrCSRBF method clearly approaches the {\abbrSSM}+\abbrRBF method over time.}
\end{figure*}

The execution time of single time and measurement updates are shown in~\cref{tab:3way:exectime}. It is clear that the proposed {\abbrSSM}+\abbrCSRBF improves over the {\abbrSSM}+\abbrRBF. On average, the time update takes approximately $\SI{0.006}{\second}$, and the measurement update roughly $\SI{0.034}{\second}$. Hence, the \abbrSSM+\abbrCSRBF filter should be able to run at a rate of approximately $\SI{25}{\hertz}$, with this particular tuning of the parameters. Further, the \abbrRMSE over number of vehicles that have passed through the intersection is visualized in \cref{fig:intersectionrmse}. The \abbrRMSE with respect to position and velocity is visualized separately for each path. As the number of vehicles increases, the \abbrRMSE both with respect to position and velocity approaches that of the {\abbrSSM}+\abbrRBF model, for both of the paths. This indicates that the proposed method does not lose any significant estimation performance (asymptotically), in the \abbrRMSE sense, at least in this particular example. Interestingly, the {\abbrSSM}+\abbrRBF method has a clear \abbrRMSE advantage already from the start. This is due to the use of an approximate Kalman gain in the {\abbrSSM}+\abbrCSRBF method, which is one of the keys to the computational complexity reduction. Thus, there is a trade-off between computational benefits and convergence rate in an \abbrRMSE sense.

\section{Conclusion and Future Work}\label{sec:conclusionandfuturework}

An extended Kalman filter method for online joint state inference and model learning has been presented and its application illustrated via two numerical examples. The underlying system dynamics are modeled using an \abbrSSM composed of a prior, known, model and an \abbrRBF expansion capturing unknown model properties. By using {\abbrRBF}s with compact support and an approximate Kalman gain in the filter it is shown that the method has a complexity of $\bigO(\tilde{n}_\weight^2)$, where $\tilde{n}_\weight$ depends on the support of the compact basis functions as well as the basis function grid density. This should be compared to the {\abbrSSM}+\abbrRBF method which has a complexity of $\bigO(n_\weight^2)$, where $n_\theta$ is the number of basis functions. Note that $\tilde{n}_\weight\leq n_\weight$ and hence, the method approaches the {\abbrSSM}+\abbrRBF formulation if the support of the basis functions is large. The drawback with the proposed method is that correlation between distant points in the state-space domain can not be represented. Still, in many applications the expected correlations are short-range and hence, the proposed method provides a computationally efficient method for joint state inference and model learning, and enables large-scale real-time applications.

Future directions concern a mathematical analysis of the model's statistical properties, \emph{i.e.}, under what conditions convergence can be guaranteed. Further, adaptive grid densities are of interest as this makes it possible to concentrate basis functions where they are needed. Lastly, methods for estimating multi-modal functions need to be addressed for this to be useful in a wider range of applications.


\begin{table}[t!]
	\centering
	{\renewcommand{\arraystretch}{1.3}%
		\caption{Three-way intersection simulation parameters\label{tab:3way:simulationparameters}}
		\subfloat[Model parameters\label{tab:3way:modelparameters}]{
			\begin{tabular}{P{.15\columnwidth}P{.2\columnwidth}m{.4\columnwidth}}\toprule
				\textbf{Parameter} & \textbf{Value} & \textbf{Description}\\\midrule
				$\delta_c$ & $\SI{1}{\meter}$ & Grid density\\
				$l$ & $1$ & \abbrRBF length scale\\
				$\alpha$ & $5$ & \abbrCSRBF support\\
				$R$ & $\SI{0.2}{\meter\per\second^2}$ & Measurement noise variance\\
				$\hat{x}_0$ & Ground truth & Prior state estimate\\
				$\Px_0$ & $0.1\cdot\vec{I}$ & Prior state error covariance\\
				$\Pw_0$ & $0.01\cdot\vec{I}$ & Prior weight error covariance\\
				$\vec{Q}$ & $0.1\vec{I}$ & Process noise covariance\\
				$\bm\Sigma$ & $\vec{0}$ & Weight noise covariance\\
				$T_s$ & $\SI{0.2}{\second}$ & Sampling interval
				\\\bottomrule
			\end{tabular}
		}\\
		\subfloat[SUMO Parameters\label{tab:3way:sumoparameters}]{
			\begin{tabular}{m{.3\columnwidth}P{.15\columnwidth}}\toprule
				\textbf{Parameter} & \textbf{Value}\\\midrule
				Vehicle length & $U(4.5, 5)$\\
				Vehicle max speed & $\SI{10}{\meter\per\second^{}}$\\
				Vehicle speed deviation & $5\%$\\
				Vehicle follow model & ACC
				\\\bottomrule
		\end{tabular}}
	}
\end{table}

\begin{table}[t!]
	\centering
	\caption{Execution times for each separate filtering step\label{tab:3way:exectime}}
	{\renewcommand{\arraystretch}{1.3}%
		\begin{tabular}{m{.2\columnwidth}P{.3\columnwidth}P{.3\columnwidth}}\toprule
			\textbf{Model} & \multicolumn{2}{c}{\textbf{Average Execution Time $[\SI{}{\second}] ~(\mathrm{std})$}}\\
			& \textbf{Time update} & \textbf{Measurement update}\\\midrule
			{\abbrSSM}+\abbrCSRBF & $0.006~(\pm 0.0027)$ & $0.034~(\pm 0.009)$\\
			{\abbrSSM}+\abbrRBF & $0.045~(\pm 0.0273)$ & $0.784~(\pm 0.065)$
			\\\bottomrule
	\end{tabular}}
\end{table}

\bibliographystyle{IEEEtran}
\bibliography{ieee-tsp2020}
\end{document}